\documentclass[%
 aip,
 amsmath,amssymb,
 reprint,%
]{revtex4-2}
\usepackage{amsfonts}

\usepackage{bm}%
\usepackage[]{hyperref}
\usepackage{graphicx}% Include figure files
\usepackage{amsmath}
\usepackage{siunitx}
\usepackage{soul}
\usepackage{amssymb}
\makeatletter
\usepackage{rotating}
\usepackage{romannum}
\usepackage{url}
\usepackage[normalem]{ulem}
\usepackage{anyfontsize}
\usepackage{blindtext}
\usepackage[utf8]{inputenc}
\usepackage{xr}
\usepackage{lipsum}
\usepackage{ccaption}

\usepackage{floatrow}
\usepackage[font= large,label font=bf,labelformat=simple]{subfig}

\usepackage[T1]{fontenc}
\usepackage{lipsum}
\usepackage[table]{xcolor} 
\usepackage{tabularx}
\begin{document}

\title{Mechanical feedback controls the emergence of dynamical memory in growing tissue monolayers}
\author{Sumit Sinha$^1$, Xin Li$^2$, Rajsekhar Das$^2$ and  D. Thirumalai$^2$}
%\email{dave.thirumalai@gmail.com}
\affiliation{Department of Physics, University of Texas at Austin, Austin, TX 78712, USA.}
\affiliation{Department of Chemistry, University of Texas at Austin, Austin, TX 78712, USA.} 

\begin{abstract}
The growth of a tissue, which depends on cell-cell interactions and biologically relevant process such as cell division and apoptosis, is regulated by a mechanical feedback mechanism. We account for these effects in a minimal two-dimensional model in order to investigate the consequences of mechanical feedback, which is controlled by a critical pressure, $p_c$.   A cell can only grow and divide if the pressure it experiences, due to interaction with its neighbors, is less than $p_c$. Because temperature is an irrelevant variable in the model, the cell dynamics is driven by self-generated active forces (SGAFs) that are created by cell division.  It is shown that even {\it in the absence of intercellular interactions}, cells undergo diffusive behavior.  The SGAF-driven diffusion is indistinguishable from the well-known dynamics of a free Brownian particle at a fixed finite temperature. When intercellular interactions are taken into account, we find persistent temporal correlations in the force-force autocorrelation function ($FAF$) that extends over timescale of several cell division times. The time-dependence of the $FAF$ reveals memory effects, which increases as $p_c$ increases. The observed non-Markovian effects emerge due to the interplay of cell division and mechanical feedback, and is inherently a non-equilibrium phenomenon.

 %A particle executing simple brownian dynamics, an example of markov process, undergoes random forcing which is delta correlated in time. However, in the presence of interactions, the properties of the emergent froces are unknown. In the present study, we performed simulations of an evolving two dimensional colony to probe the the role of mechanical feedback ($p_c$). We calculated the force-force auto-correlation (FAF) on individual cells and discovered the emergence of persistent correlations extending over several cell division timescale. The magnitude of correlations increases on increasing $p_c$. The emergence of persistent FAF shows that the dynamics of individual cells provides might indicate non-markovian/memory effects.

\end{abstract}


%\begin{titlingpage}
\maketitle

\section{Introduction}
\label{introduction}
Life around us, spanning a bewildering array of length and time scales, is sustained through multicellular processes that are driven by non-equilibrium events such as cell growth and cell division \cite{alert2020physical,shaebani2020computational}. Although known for a long time \cite{Hemlinger97NatBiotech}, several recent experimental studies have emphasized that growth and division in cell collectives are governed by local stresses that the cells experience\cite{delarue2016self, Dolega17NC, puliafito2012collective, shraiman2005mechanical}. A manifestation of coupling of growth and division to local stress is the deviation of growth law of the cell collective from exponential law \cite{puliafito2012collective}. These experiments suggest that there must exist mechanical feedback between the local stress and cell division. 

In a series of papers, we showed that single cells in a collective exhibit anomalous dynamics due to local stress-dependent cell growth and division \cite{malmi2018cell, malmi2019dual, sinha2020self, sinha2020spatially, samanta2020far, sinha2021inter, samanta2019origin}. In the present study, we explore the dependence of mechanical feedback, mediated by stress threshold $p_c$, on the dynamics of single cells. A recent interesting study \cite{gniewek2019biomechanical}, has shown that mechanical feedback regulates the physical properties of jammed cell collectives, which supports experimental findings \cite{delarue2016self}. However, how the mechanical feedback regulates individual cell migration in a collective is unknown, and is the problem which we address in this study. 

Using a two-dimensional off-lattice agent-based simulation model, we explore the role of mechanical feedback ($p_c$) on single-cell dynamics. The central results of the present study are: (\romannum{1}) In the absence of cell growth and division, the dynamics is solely governed by short-ranged two body interactions. In this limit, the cells in the long-time behave like a glass-like solid. (\romannum{2}) In the presence of cell division, with {\it systematic interactions absent}, the cells exhibit Markov dynamics resulting in diffusive motion at long times. This finding is surprising because there is no thermal motion (temperature is an irrelevant variable). This is different from a free Brownian particle where temperature randomizes the particle motion. (\romannum{3})  When both cell division and systematic interactions control the collective movement  in tandem, the dynamics is regulated by the mechanical feedback that is parameterized using $p_c$. To quantify the dynamics, we calculated the force auto-correlation function (FAF) inspired by works in the theory of chemical reactions in liquids \cite{straub1987calculation,straub88JCP,berne1970correlation}. We show FAF  increases when $p_c$ is increased. The emergence of long time correlation in the FAF shows  departure from Markov dynamics, and is suggestive of memory effects in growing cell collectives. (\romannum{4}) The persistence of trajectories of individual cells increases as $p_c$ is increased.  The trajectories are strikingly different from simple Brownian motion. The enhanced persistence in cell dynamics, as $p_c$ increases, is the origin of memory in active systems. Taken together, the present study establishes how mechanical feedback coupled with cell growth and division leads to non-Markovian cell dynamics whose importance has not been appreciated before. 

\begin{figure*}[h]
\centering 
\includegraphics[width=15cm]{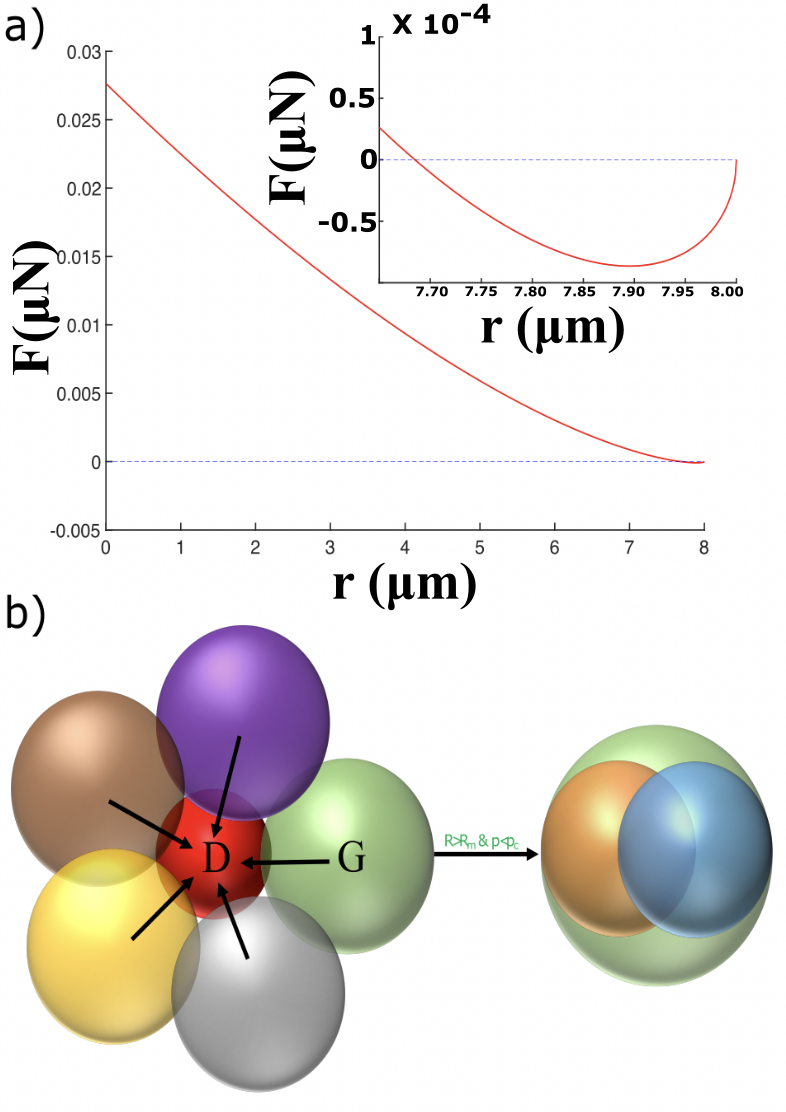}
\caption{{\bf Schematic of the 2D model}.{\bf (a)} Total force (Eq. \ref{total}) as a function of inter-cellular distance for two cells, $i$ and $j$, with radius $R_i = R_j = 4~\mu m$. The repulsive and attractive part of the force are given by Eqs. \ref{rep} and \ref{ad}, respectively. The inset is the zoomed in view that highlights the region in which the force is predominantly attractive.  {\bf (b)} Cartoon illustrating the role of mechanical feedback. On the left, the `red' cell is dormant (cannot grow and divide) because the pressure exerted by the neighbors exceeds $p_c$. The `green' cell is in the growth phase ($G$), which grows ands divides ($p<p_c$). The green cell from the left gives given birth to two daughter cells (orange and cyan) when the radius exceeds the mitotic radius $R_m$.      }
\label{cartoon}
\end{figure*}

\section{Methods}
\label{methods}
We simulated the spatial and temporal dynamics of a two dimensional (2D) growing tissue using agent-based off-lattice model in which the cells are represented as interacting deformable disks. This simplified assumption of representing cells as deformable disks was also used in previous studies \cite{matoz2017cell}, although the details differ. In the model, the cells grow stochastically in time and divide upon reaching a critical size ($R_m$), the mitotic radius. The cell-to-cell interaction is characterized by elastic and adhesive forces. We also consider cell-to-substrate damping as a way of accounting for the effects of friction experienced by a moving cell by the substrate.

%\textbf{Details of the model:}
{\bf Physical Interactions :} Each cell is modeled as a deformable disk with a time dependent radius. 
A cell is characterized by physical properties such as the radius, elastic modulus, membrane receptor and ligand 
concentration. In addition, the cells attract each other through E-Cadherin mediated adhesive interactions. This model is inspired by previous works on 3D off-lattice multicellular tumor growth models \cite{drasdo2005single, schaller2005multicellular, malmi2018cell, malmi2019dual, sinha2020spatially, samanta2020far, sinha2020self, sinha2021inter}. The elastic (repulsive) force between two disks with radii $R_{i}$ and $R_{j}$ is given by,
\begin{equation}
\label{rep}
F_{ij}^{el}(t) = \frac{h_{ij}^{3/2}(t)}{\frac{3}{4}(\frac{1-\nu_{i}^2}{E_i} + \frac{1-\nu_{j}^2}{E_j})\sqrt{\frac{1}{R_{i}(t)}+ \frac{1}{R_{j}(t)}}},
\end{equation}
where $E_{i}$ and $\nu_{i}$, respectively, are the elastic modulus and
Poisson ratio of cell $i$. The overlap between the disks, if they interpenetrate without deformation, is $h_{ij}$, which 
is given by $\mathrm{max}[0, R_i + R_j - |\vec{r}_i - \vec{r}_j|]$ with $|\vec{r}_i - \vec{r}_j|$ being the center-to-center distance between the two disks. 
%\floatsetup[figure]{style=plain,subcapbesideposition=top}
%\begin{figure}
%\sidesubfloat[]{\includegraphics[width=0.7\linewidth] {cellcellinter.eps}\label{cellcellinter}}
%	\par\bigskip
%\sidesubfloat[]{\includegraphics[width=0.70\linewidth] {forcejkr.eps}\label{forcejkr}} 
%\caption{\textbf{(a)} Illustration of two interpenetrating cells $i$ and $j$ with radii $R_{i}$ and $R_{j}$, respectively.  The distance between the centers of the two cells is $|{\mathbf {r}}_{i} -{\mathbf {r}}_{j}|$, and their overlap is $h_{ij}$.
%\textbf{(b)} Force on cell $i$ due to $j$, ${\mathbf F}_{ij}$, for $R_{i}=R_{j}=4~\mu m$ using mean values of elastic modulus, poisson ratio, receptor and ligand concentration (see Table I). ${\mathbf F}_{ij}$ is plotted 
%as a function of distance between the centers of the two cells. 
%Inset shows the region where ${\mathbf F}_{ij}$ is attractive. When $|{\mathbf {r}}_{i} -{\mathbf {r}}_{j}|~\ge~R_{i}+R_{j}=8~\mu m$ the cells are no longer in contact, and hence, ${\mathbf F}_{ij}=0$.}
%end{figure}

Cell adhesion, mediated by receptors on the cell membrane,  
is the process by which cells can attach to one another. 
For simplicity, we assume that the receptor and ligand molecules are evenly 
distributed on the cell surface. Consequently, the magnitude of the adhesive force, $F_{ij}^{ad}$, 
between two cells $i$ and $j$ is expected to scale as a function of their contact line-segment, 
$L_{ij}$. Keeping the 3D model as a guide \cite{malmi2018cell}, we calculate $F_{ij}^{ad}$ using,
\begin{equation}
\label{ad}
F_{ij}^{ad} = L_{ij}f^{ad}\frac{1}{2}(c_{i}^{rec}c_{j}^{lig} + c_{j}^{rec}c_{i}^{lig}),
\end{equation}
where the $c_{i}^{rec}$ ($c_{i}^{lig}$) is the receptor (ligand) concentration 
(assumed to be normalized with respect to the maximum receptor or ligand concentration so that  
$0 \leq c_{i}^{rec},  c_{i}^{lig} \leq 1$). The coupling constant $f^{ad}$ allows us to 
%\leq c_{i}^{(rec/lig)/max} 
rescale the adhesion force to account for the variabilities in the maximum densities of the receptor and ligand concentrations.  
We calculate the contact length, $L_{ij}$, using the length of contact between two intersecting circles,  $L_{ij} = \frac{\sqrt{(|4r_{ij}^2R_i^2-(r_{ij}^2-R_j^2+R_i^2)^2|)}}{r_{ij}}$. Here, $r_{ij}$ is the distance between cells $i$ and $j$. As before, $R_i$ and $R_j$ denote the radius of cell $i$ and $j$. 

Repulsive and adhesive forces considered in Eqs.(\ref{rep}) and (\ref{ad}) 
%act on the center of the spheres 
act along the unit vector ${\bf n}_{ij}$ pointing from the center of cell $j$ to the center of cell $i$. 
%The force exerted by cell $j$ on cell $i$, ${\bf F}_{ij}$, is shown in figure ~\ref{force_distance}, with the positive part denoting the repulsive regime and negative part depicting the attractive regime.
The total force on the $i^{th}$ cell is given by the sum over its nearest neighbors ($NN(i)$), 
\begin{equation}
{\bf F}_{i} = \Sigma_{j \epsilon NN(i)}(F_{ij}^{el}-F_{ij}^{ad}){\bf n}_{ij}. 
\label{total}
\end{equation}
The nearest neighbors satisfy the condition $R_i + R_j - |{\bf r}_i - {\bf r}_j|~>~0$.
Figure \ref{cartoon}a shows the plot of the total force as a function of inter-cellular distance.

{\bf Three Scenarios:} In order to elucidate the dramatically different dynamical behavior, we consider three limits. (I) The collective movement arising solely from the systematic forces, given in Eq. \ref{total}. (II) Cell movement with ${\bf F}_{i}=0$ (no inter-cellular interactions) but 
allowing for cell division and growth. Note that since inter-cellular interactions are absent, mechanical feedback ($p_c$) does not play a role. In this limit, we show that the dynamics can only arise due to active forces generated upon cell division. The limits (I) and (II) are not relevant in describing collective movements in Multicellular Spheroids (MCSs) \cite{valencia2015collective} or evolving cell monolayers \cite{puliafito2012collective}.   (III) In this limit, we not only include interactions between cells (Eq. \ref{total}) but also allow for cell growth, division and apoptosis. Most importantly, the time-dependent growth of the tissue colony is limited by mechanical feedback, which prohibits the biologically important process of cell growth and division if the local stress on a cell exceeds a critical non-zero value, $p_c$.

%{\bf SS: We should have a figure showing total force as a function of r and the case of dormancy. (a) and (b).}
%\section{Simulations}

%\textbf{Equations of Motion:} %Once the force is calculated, 

{\bf Equation of Motion:} The damped dynamics of the $i^{th}$ cell is computed based 
on the equation of motion, 
\begin{equation}
\label{eqforce}
\dot{{\bf r}}_{i} = \frac{{\bf F}_{i}}{\gamma_i}.
\end{equation}
Here, $\gamma_i$ is the friction coefficient of the $i^{th}$ cell. We assume $\gamma_i$ to be equal to $\gamma_o R_i(t)$, where $\gamma_o$ is a constant. This form of $\gamma_i$ is inspired from the simulations of three dimensional models for solid tumor where $\gamma_i=6\pi \eta R_i$ with $\eta$ being the viscosity. %Therefore, $\gamma_o=6 \pi \eta$ for the three dimensional case. 
Note, we do not consider the effect of temperature (set to zero in the simulations) as we assume the friction coefficient, that in reality arises from the extracellular matrix in 3D or substrate in 2D, to be so high \cite{matoz2017cell} that thermal motion is irrelevant. The equation of motion in Eq. \ref{eqforce} is similar to the case for soft granular materials where the role of temperature is neglected \cite{chacko2019slow}. However, it is crucial to note that in the growth of the tissue colony, scenarios II and III in our case, there is a self-generated active force (SGAF) that arises due to the biologically important processes of cell growth and division \cite{sinha2020self}.

% \begin{figure}
%\begin{turn}{-90}
%\includegraphics[width=0.90\linewidth] {celldorgro.eps} %oa
%\end{turn}
%\caption{Cell dormancy (left panel) and cell division (right panel). If the local pressure $p_i$ that the $i^{th}$  cell experiences 
%(due to contacts with the neighboring cells) exceeds the critical pressure $p_{c}$, it enters 
%the dormant state ($D$).  Otherwise, the cells grow (G) until they reach the mitotic radius, $R_{m}$. 
%At that stage, the mother cell divides into two identical daughter cells with the same radius $R_{d}$. 
%We assume that the total volume upon cell division is conserved. A cell that is dormant at a given time can transit from that state at subsequent times.}
%\label{celldorgro}
%\end{figure}
{\bf Cell growth, division and apoptosis:}
In our model, cells can be either in the dormant ($D$) or in the growth ($G$) phase depending on the local pressure associated with a cell (Figure \ref{cartoon}(b)).
Using Irving-Kirkwood definition, we track the pressure ($p_i$) experienced by the $i^{th}$ cell due to contact with its neighbors \cite{yang2014aggregation}. The expression for  $p_i$ is given by,
\begin{equation}
\label{pressure}
p_{i} =  \frac{1}{2}\Sigma_{j \epsilon NN(i)} \frac{{\bf F}_{ij} \cdot {\bf dr}_{ij}} {A_i},
\end{equation}
where $A_i = \pi R_i^2$, is the area of the cell.  If the local pressure, $p_{i}$, exceeds a critical limit ($p_c$) the cell stops growing and enters the dormant phase. Note, the cell can switch back to a growing phase if $\frac{p_i}{p_c}<1$ as the tissue evolves. The critical pressure $p_c$, serves as a mechanical feedback, which is known to regulate the growth of tissues \cite{shraiman2005mechanical}. 
%The cell is considered to be in the dormant state. 

For growing cells ($\frac{p_i}{p_c}<1$), their area increases at a constant rate $r_A$. 
The cell radius is updated from a Gaussian distribution with the mean rate $\dot{R} = (2\pi R)^{-1} r_A$.  
Over the cell cycle time $\tau$, 
\begin{equation}
r_A = \frac{\pi (R_{m})^2}{2\tau},
\end{equation}
where $R_{m}$ is the mitotic radius. The cell cycle time ($\tau$) is related to the growth rate ($k_b$) by $\tau=\frac{ln~2}{k_b}$. 
A cell divides once it grows to the fixed mitotic radius ($R_m$). 
To ensure area conservation, upon cell division, we use 
$R_d = R_{m}2^{-1/2}$ as the radius of the daughter cells. The two resulting cells are placed at a center-to-center distance 
$d = 2R_{m}(1-2^{-1/2})$. The direction of the new cell location 
is chosen randomly from a uniform distribution on the unit circle. 
One source of stochasticity in the cell movement in our model is due to random choice for  
the mitotic direction. In our simulations, the cells may undergo apoptosis at the rate $k_a$. Throughout this work, the apoptosis rate was fixed to $10^{-6} s^{-1}$. Table \Romannum{1} depicts the parameters used in the simulations.

\begin{table}[ht]
\centering 
\caption{The parameters used in the simulation.}
\scalebox{1}{\begin{tabular}{ |p{4cm}||p{2cm}|p{2cm}|p{2cm}|  }
 \hline
 \bf{Parameters} & \bf{Values} & \bf{References} \\
 \hline
 Timestep ($\Delta t$)& 10$\mathrm{s}$  & This paper \\
 \hline
Critical Radius for Division ($R_{m}$) &  5 $\mathrm{\mu m}$ & ~\cite{schaller2005multicellular, malmi2018cell}\\
 \hline
Friction coefficient ($\gamma_o$) & 0.1 $\mathrm{kg/ (\mu m~s)}$   & This paper  \\
 \hline
 Cell Cycle Time ($\tau$)  & 54000 $\mathrm{s}$  & ~\cite{freyer1986regulation, casciari1992variations,landry1981shedding,malmi2018cell}\\
 \hline
 Adhesive Coefficient ($f^{ad})$&  $10^{-4} \mathrm{\mu N/\mu m}$  & This paper \\
 \hline
Mean Cell Elastic Modulus ($E_{i}) $ & $10^{-3} \mathrm{MPa}$  & ~\cite{galle2005modeling,malmi2018cell}    \\
 \hline
Mean Cell Poisson Ratio ($\nu_{i}$) & 0.5 & ~\cite{schaller2005multicellular,malmi2018cell}  \\
 \hline
 Apoptosis Rate ($k_a$) & $10^{-6} \mathrm{s^{-1}}$ & ~\cite{malmi2018cell} \\
 \hline
Mean Receptor Concentration ($c^{rec}$) & 1.0 & ~\cite{malmi2018cell} \\
\hline
Mean Ligand Concentration ($c^{lig}$) & 1.0  & ~\cite{malmi2018cell}  \\
\hline
%\label{table1}
%\caption{Parameters used to carry out the simulations.}
\end{tabular}}
\end{table}

\textbf{Initial Conditions:} 
We begin the simulations by placing 100 cells on a 2D plane whose coordinates are chosen from a normal distribution with  mean zero and standard deviation $25~\mu m$. All the parameters apart from critical pressure ($p_c$) are fixed. 

\begin{figure*}[h]

\includegraphics[width=18cm]{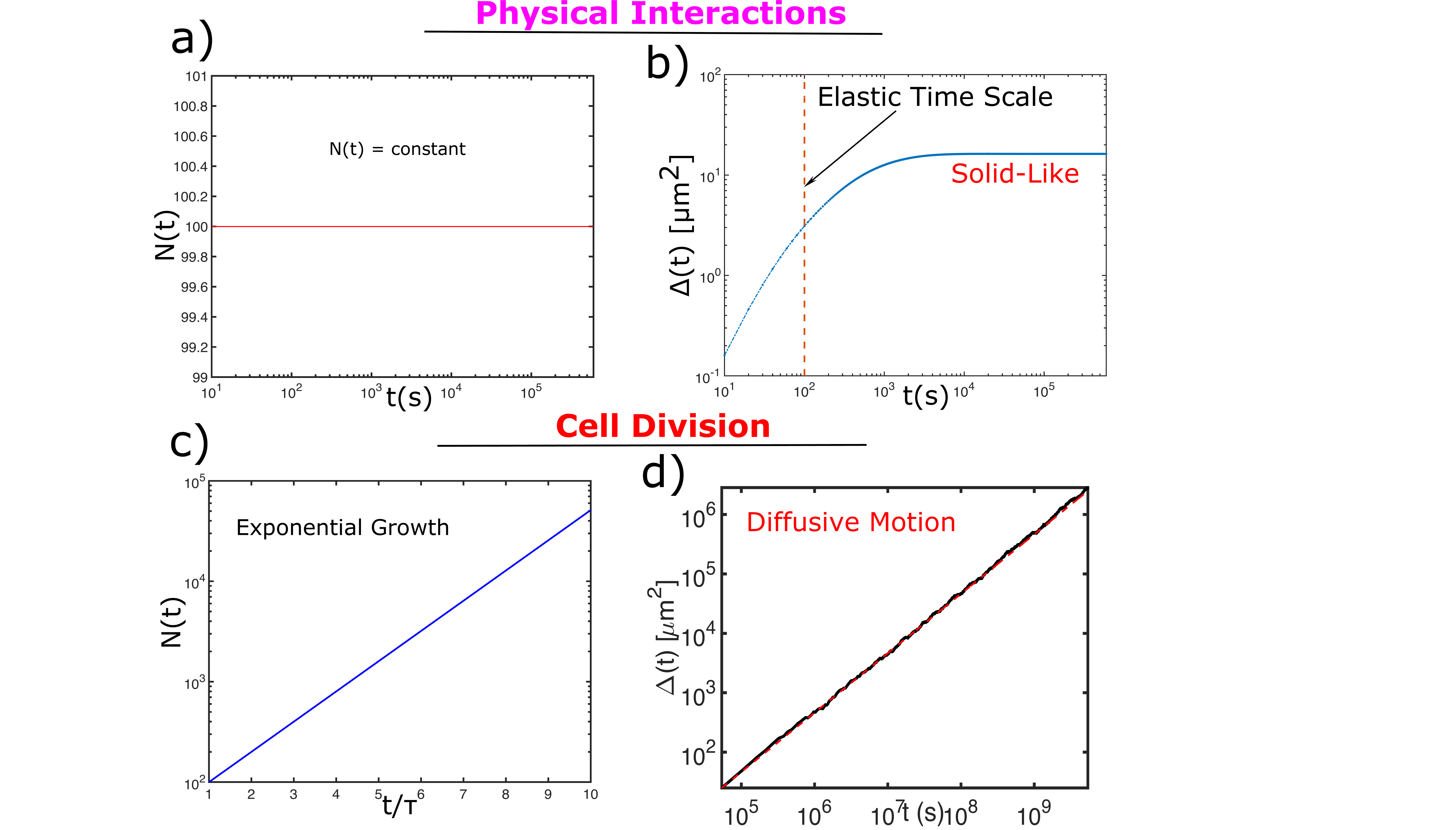}
\caption{\textbf{Number and MSD of cells solely based on physical interactions and cell division:} {\bf (a)} Number of cells, $N(t)$, solely based on physical two body interactions. Since, cell division rate is zero, $N(t)$ is a constant. {\bf (b)} Mean-Squared Displacement, $\Delta(t)$, as a function of time when only physical interactions are present. $\Delta(t)$ relaxes initially, governed by elastic forces over a time-scale $\frac{\gamma}{ER_m}$ (red dashed vertical line), and settles to a plateau value. {\bf (c)} $N(t)$ as a function of scaled time ($\frac{t}{\tau}$) in the presence of cell division only. $N(t)$ grows exponentially. {\bf (d)} $\Delta(t)$ as a function of time in the presence of cell division only. $\Delta(t)$ grows linearly in time and is diffusive (in black). The red dashed line is a linear fit given by Eq. \ref{cell_diff}.}
\label{toy}
\end{figure*}

\section{Results}

\subsection{Markov dynamics in the presence of delta-correlated random force}
For comparison, we briefly summarize the well-known result for a  stochastic process (over-damped Langevin equation) in one dimension for a free Brownian particle. The equation of motion is, 
\begin{equation}
    \frac{dx}{dt}=\sqrt{2D}\eta(t),
    \label{brownian_mot}
\end{equation}
where the random force obeys $\langle \eta(t) \rangle=0$ and  $\langle \eta(t)\eta(t') \rangle=\delta(t-t')$. The position of the particle is  $x$. The solution $P(x,t)$, for probability density for finding the particle at $x$ at time $t$, is given by 
\begin{equation}
P(x,t)=\frac{1}{\sqrt{4\pi D t}}e^{-\frac{x^2}{4Dt}}.
\end{equation}
Here, we have assumed that the particle was  at the origin at $t=0$, $P(x,0)=\delta(x)$. The moments of $P(x,t)$, which serve as the physical observables in cell tracking experiments \cite{valencia2015collective}, are readily calculated. For instance, the first moment $\langle x \rangle$ is given as 
\begin{equation}
\langle x(t)\rangle= \int_{-\infty}^{\infty}xP(x,t)=0.
\end{equation}
The second moment $\langle x^2 \rangle$, also called the mean-squared displacement (MSD), is non-zero and is given as 
\begin{equation}
\langle x^2(t) \rangle=\int_{-\infty}^{\infty}x^2P(x,t)=2Dt.
\label{brownian_mot_msd}
\end{equation}
The dynamics of a free Brownian particle is an example of a Markov process. There is no memory because $P(\delta x(t))=\mathcal{N}(0,\sqrt{2D\delta t})$ is independent of $x(t)$, where $\delta x(t)=x(t+\delta t)-x(t)$ is the displacement of the particle from time $t$ to $t+\delta t$ and $\mathcal{N}(0,\sqrt{2D\delta t})$ is the normal distribution with zero mean and variance $2D\delta t$. This example sets the stage for exploring the emergent force auto-correlation ($FAF$), with long temporal correlation,  in an expanding tissue.

\subsection{Role of physical interactions and cell division}
We first investigate the consequences of physical interactions and cell division when they are not coupled to one another. 

(a) {\bf Physical Interactions (limiting case I):} When the interactions between cells are based only on the systematic interactions, as given in Eqs. \ref{rep} and \ref{ad} without cell division and apoptosis,  the dynamics of the interacting cells is governed by the elastic timescale $\frac{\gamma}{ER_m}$. In the absence of cell growth, division and apoptosis, the number of cells, $N(t)$, is a constant, which is confirmed in Figure \ref{toy}a. Figure \ref{toy}b shows the plot of mean-square displacement, $\Delta(t)$, defined as 
\begin{equation}
    \Delta(t)=\bigg\langle \frac{1}{N}\sum_{i=1}^{N} \big({\bf r}_i(t)-{\bf r}_i(0)\big)^2 \bigg \rangle,
\end{equation}
where $\langle ... \rangle$ denotes the ensemble average over $20$ simulation runs and $N$ is the initial $100$ cells. Figure \ref{toy}a shows that $\Delta(t)$ relaxes rapidly to a plateau on a time scale $\approx \frac{\gamma}{ER_m}$. Usually, $\Delta(t)\sim t^{\alpha}$, with $\alpha=0$, is indicative of solid-like behavior. Figure \ref{toy}b shows that in the long time limit, $t>>\frac{\gamma}{ER_m}$, $\alpha=0$, and hence the cell collective behaves as a solid in the sense there is absence of diffusion.  Note, the dynamics is performed under athermal (temperature is not relevant) open boundary conditions and the scale of systematic interactions are short-ranged ($\approx R_m$). Hence, the cells cannot move after the initial relaxation process. 

(b) {\bf Effect of cell division (limiting case II):} In this limit, during each cell division event, a cell is displaced by the distance $\approx R_m$, randomly in space. Hence, when the time evolution of cell colony is governed solely by cell division (absence of apoptosis and systematic interactions) we expect that with successive  cell divisions a cell would undergo a random walk that is uncorrelated in time and space. In other words, it would behave as a Brownian particle due to the SGAF induced by cell division. This is purely a non-equilibrium dynamical process. As a result, we expect that $\Delta (t)=D_{eff} t$ at long times. Because in this limiting case $\tau$ is the only time scale and $R_m$ is the only length scale, we obtain $D_{eff} = \frac{R_m^2}{\tau}$.  In the absence of systematic interactions the pressure (Eqn. \ref{pressure}) on an individual cell is zero, all the cells grow and divide independently. Hence, the number of cells, $N(t)$ increases exponentially (see Figure \ref{toy}c). In this limit, 
\begin{equation}
\frac{dN(t)}{dt}=k_b N(t)
\end{equation}
and therefore $N(t)=N_oe^{k_bt}$. Interestingly, in accord with the arguments given above, the dynamics of individual cells is diffusive in this limit and the mean squared displacement $\Delta(t)$ is given by
\begin{equation}
 \Delta(t)=\frac{R_m^2}{\tau}t,
 \label{cell_diff}
\end{equation}
as shown in Figure \ref{toy}d. Although there is no thermal motion, the scaling behavior of $\Delta(t)$, is similar to the standard Brownian dynamics given by Eqn. \ref{brownian_mot_msd}.  The dynamics is diffusive because during every cell division the cell is displaced by distance $R_m$ randomly, and hence mimics a Brownian motion in two dimensions. Note, when only cell apoptosis is present ($k_b=0$), the cells do not move,  and $N(t)$ decreases exponentially to zero at the rate $k_a$. 

The dynamics is non-trivial when all the components (systematic forces, cell division and apoptosis, and mechanical feedback)  are included. For this case, the growing tissue develops a core where the cells are jammed, and exhibit glass-like dynamics \cite{sinha2020spatially}. In contrast, the cells in the periphery are predominantly in the growth ($G$) phase. As a result, the cells exhibit anomalous spatially heterogeneous dynamics with super-diffusive (sub-diffusive) periphery (core) \cite{sinha2020spatially}. Our previous work has shown that this dynamical phase separation arises due to SGAFs that arise due to local stress-regulated cell growth and division \cite{sinha2020self}. In the model, cell division and growth are regulated by the mechanical feedback parameter $p_c$, the consequences of which is explored in the following section. 

\begin{figure*}[h]
\centering 
\includegraphics[width=19cm]{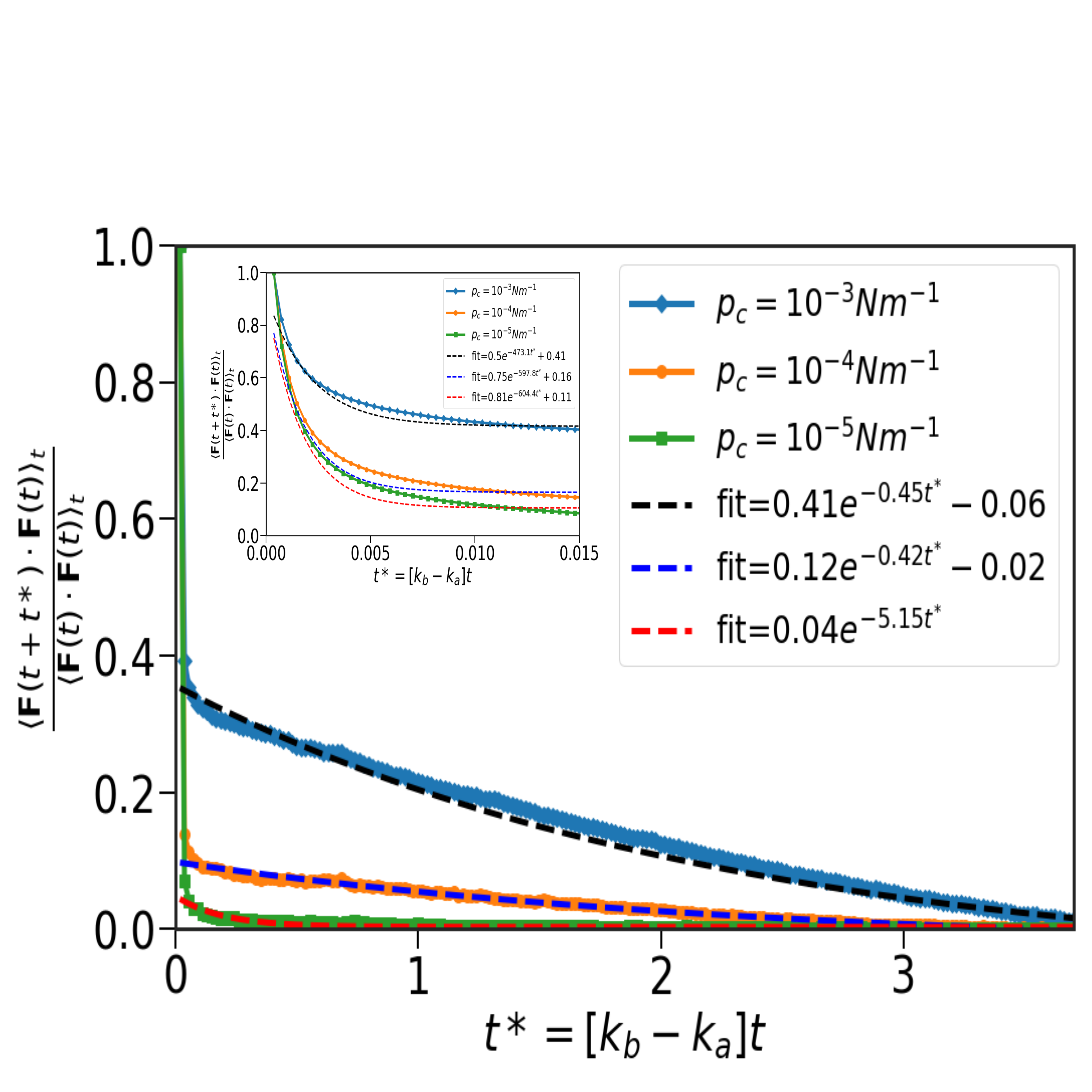}
\caption{\textbf{Emergence of highly correlated force:} Plot of force auto-correlation function (FAF) as a function of time. From top to bottom, FAF corresponds to $p_c=10^{-3}, 10^{-4}$ and $10^{-5}$. The dashed line corresponds to the fits. In the inset, we zoom in on the initial time regime of FAF. The order of the plots and dashed lines are same as in the main figure. The figure shows the emergence of FAF with two time scales: one long ($\sim \frac{1}{k_b-k_a}$) and one small (elastic time scale $=\frac{\gamma}{ER_m}$).   }
\label{force_auto}
\end{figure*}

\subsection{Highly correlated force correlations in an expanding tissue}
\label{emerge}
For a simple Brownian motion, the force is delta-correlated and hence the dynamics is Markovian (see Eqn. \ref{brownian_mot}). Therefore, a signature of such dynamics is the fast decay of force auto-correlation function ($FAF$), in comparison to the smallest time-scale of the problem, which in the present case is  $\frac{\gamma}{ER_m}$. To explore the nature of the dynamics, when both cell division and systematic interactions are present in tandem, we calculated $FAF$ ($t^{*}$) given as
\begin{equation}
FAF(t^{*})=\frac{\langle{\bf F}(t+t^{*})\cdot {\bf F}(t)\rangle_t}{\langle{\bf F}(t)\cdot {\bf F}(t)\rangle_t}
\end{equation} 
Here, ${\bf F}(t)$ is the force on the cell at time $t$ and $\langle...\rangle_t$ is the time average, and $t^{*}$ is the delay (or waiting) time. The time averaging is performed over $\approx 2,000$ cells. Figure \ref{force_auto} shows the plot of FAF for $p_c=10^{-3} Nm^{-1}, 10^{-4} Nm^{-1}$ and $10^{-5} Nm^{-1}$. Figure \ref{force_auto} shows that the FAF decays on two  time scales: long ($\sim \frac{1}{k_b-k_a}$) and  short ($\frac{\gamma}{ER_m}$). In order to extract the two time scales, we fit FAF with $Ae^{\frac{-t^{*}}{\tau_c}}+B$ in both the regimes. 

In the short time regime (see the inset of Figure \ref{force_auto}), for $p_c=10^{-3} Nm^{-1}$, $A=0.5, \tau_c=\frac{1.2\gamma}{ER_m}$ and $B=0.41$. For $p_c=10^{-4} Nm^{-1}$, $A=0.75, \tau_c=\frac{0.97\gamma}{ER_m}$ and $B=0.16$. Lastly, for $p_c=10^{-5} Nm^{-1}$, $A=0.81, \tau_c=\frac{0.95\gamma}{ER_m}$ and $B=0.11$. As anticipated, we find that in the short time regime, the relaxation time is approximately close to the elastic time scale $\frac{\gamma}{ER_m}$, which is negligible compared to $\frac{1}{k_b-k_a}$.
However, in the long time limit,  FAF shows memory effects, especially  for $p_c=10^{-3} Nm^{-1}$. For $p_c=10^{-3} Nm^{-1}$, $A=0.41, \tau_c=\frac{2.2}{k_b-k_a}$ and $C=-0.06$. For $p_c=10^{-4}$, $A=0.12, \tau_c=\frac{2.3}{k_b-k_a}$ and $B=-0.02$. Lastly, for $p_c=10^{-5} Nm^{-1}$, $A=0.04, \tau_c=\frac{0.2}{k_b-k_a}$ and $B\approx 0$. For $p_c=10^{-5} Nm^{-1}$, A is negligible which indicates the absence of memory. In the other two cases, A for $p_c=10^{-3} Nm^{-1}$ is four times larger than for $p_c=10^{-4} Nm^{-1}$. Larger magnitude of FAF in the long time regime leads to higher degree of migration for $p_c=10^{-3} Nm^{-1}$.
The emergence of highly correlated forces elucidates the departure from Markovian dynamics in a system comprising of many interacting cells whose time evolution occurs under non-equilibrium conditions, and absence of fluctuation-dissipation theorem \cite{samanta2019origin}.

\begin{figure*}[h]
\centering 
\includegraphics[width=20cm]{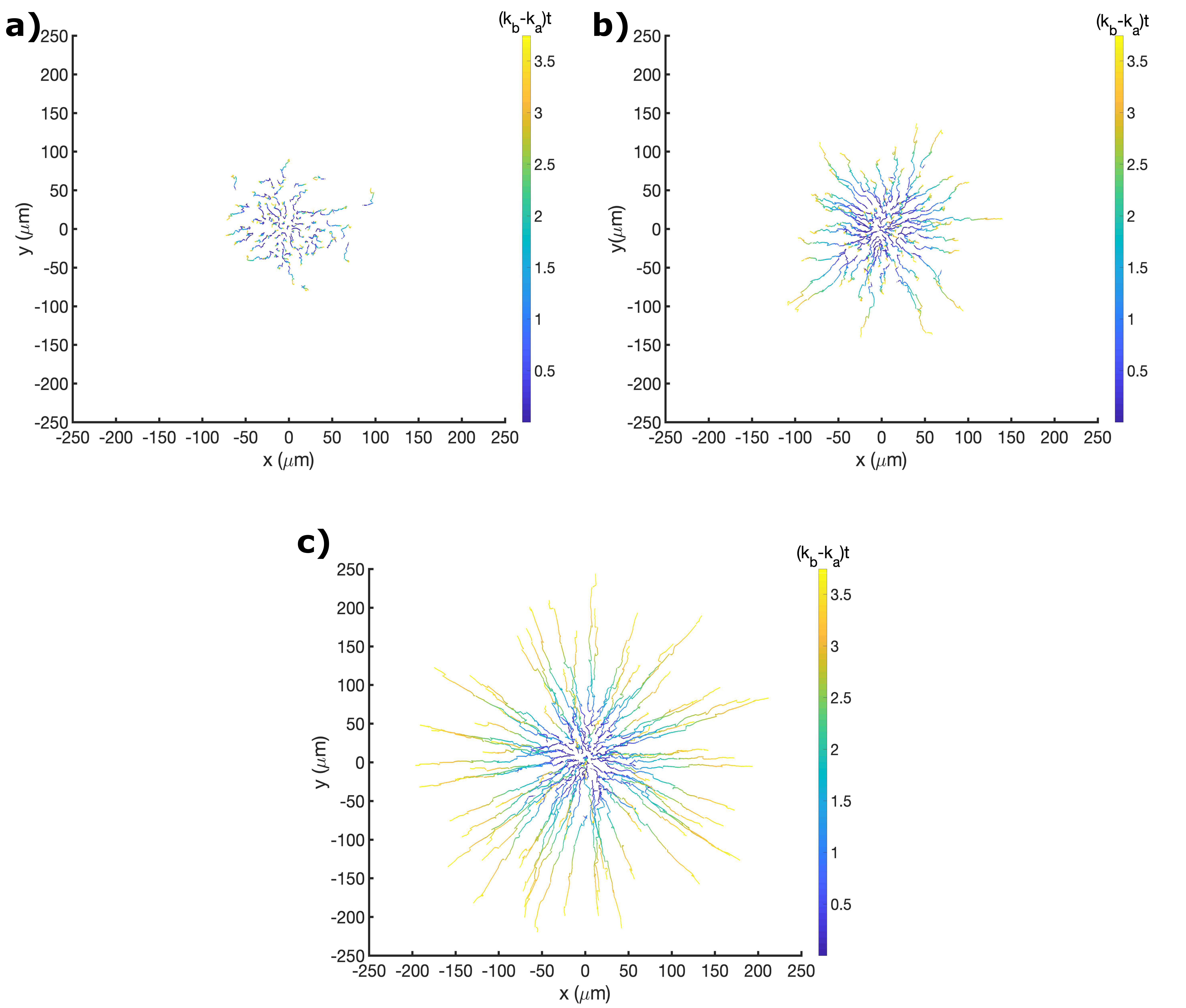}
\caption{\textbf{Trajectories of cells for different $p_c$ values:} {\bf (a, b, c)} Trajectories of cells for $p_c=10^{-5}~N/m$, $p_c=10^{-4}~N/m$ and $p_c=10^{-3}~N/m$. For all the three panels, the color bar represents the time in units of $(k_b-k_a)t$. From the figures, it is clear that cells migration is enhanced when $p_c$ is increased.    }
\label{cell_traj}
\end{figure*}

\subsection{Persistence of trajectories increases on increasing $p_c$ }
\label{persistence}
To better understand, the significance of memory effects that are embodied in FAF, we probed the trajectories of individual cells in the growing tissue colony. We investigated the three cases: $p_c= 10^{-5}~N/m, 10^{-4}~ N/m$ and $10^{-3}~N/m$ in Figure \ref{cell_traj}. Note, the trajectories are for  the initial $100$ cells with the identical initial conditions  and the total simulation time.  Thus, the differences in the trajectories emerge due to interplay of systematic interactions, cell growth and division. Figure \ref{cell_traj} shows that the nature of the trajectories are strikingly different when $p_c$ values are changed. For $p_c=10^{-5}~N/m$ in Figure \ref{cell_traj}a, the cells move the smallest compared to $p_c=10^{-4}~N/m$ and $p_c=10^{-3}~N/m$ and are not persistent in nature. This is because the the memory effect in the $FAF$ is negligible, as illustrated in Figure \ref{force_auto}. The cells exhibit persistent trajectories for  $p_c=10^{-4}~N/m$ and $p_c=10^{-3}~N/m$, with the degree of persistence being higher for the latter. The emergence of persistence in trajectories implies the presence of memory or (non-Markovian) effects in the dynamics.  
\section{Conclusion}
We have shown that in an evolving tissue colony, the dynamics of an individual cell may be approximated as a stochastic process where the force is correlated over many cell division times.  The emergence of correlation in force is a manifestation of memory effects, which is a hallmark of non-Markovian dynamics. Memory effects in a growing tissue arises due to cell division and mechanical feedback. There is no potential or energy function whose derivative results in these processes. This immediately implies that there is no equilibrium in the growing tissue, and hence the system is always out of equilibrium.
The present study also provides a mechanism for persistent motion observed in many out of equilibrium active matter systems \cite{gompper2020, bechinger2016active,Marchetti13RMP}. 

It is tempting to describe the simulated time-dependent force autocorrelation using a reduced description, something like the Generalized Langevin equation (GLE), to describe the effective dynamics of a cell in the evolving tissue. It is natural to use such an approach in thermally controlled barrier crossing problems, as was done decades ago  in the most insightful studies \cite{straub1987calculation,straub88JCP}.   For reasons stated above, construction of a similar set of equations, if it exists at all, in any reduced variable (the analogue of reaction coordinate in barrier crossing problems), which must also include large spatial heterogeneity, is likely to be difficult.  Cell-division and apoptosis require energy input and dissipation, which can only be described using a physical picture and a framework that goes beyond the usual description based on Hamiltonians or energy functions.  The work here may provide an impetus to develop  a general theoretical framework for describing feedback controlled dynamics in active systems.

\section*{Acknowledgement}
We would like to thank Abdul N. Malmi-Kakkada and Himadri S. Samanta for valuable comments on the manuscript. This work was supported by grants from National Science Foundation (PHY 17-08128, PHY-1522550). Additional support was provided by the Collie-Welch Reagents Chair (F-0019). 

\section*{DATA AVAILABILITY}
The data that support the findings of this study are available from the corresponding author upon reasonable request.

%\clearpage

%\clearpage

%\clearpage

%\clearpage

\vskip 0.2in
\bibliographystyle{unsrt}
\bibliography{memory}

\end{document}